# Magnetic properties of quasi one-dimensional vanadium-benzene nanowire affected by gas molecules: a first-principle study

Min Wang, Yan Zhou, Sui Kong Hark , Xi Zhu

School of Science and Engineering,

The Chinese University of Hong Kong,Shenzhen,

Shenzhen, Guangdong, 518172

## Abstract

Magnetic properties of quasi one-dimensional vanadium-benzene nanowires (VBNW) are investigated theoretically with the absorption of gas molecules-NO and $NO_2$. With the increase adsorption of NO on VBNW, the phase transition from half metal to ferromagnetic metal and last to paramagnetic semiconductor can be observed. With the increase of $NO_2$ on VBNW, half metallic property can be enhanced at first and decreased later. Thus, the electronic and magnetic properties of VBNW can be sensitive and selective to NO and $NO_2$, revealing the potential applications in spintronic sensors of these two kinds of molecules.

## 1. Introduction

Different from traditional spintronics, molecular spintronics provide new challenges in fundamental understandings about the charge and spin transfer mechanism and also hold new opportunities in potential applications of future electronic device[1-3]. Among all the molecular spintronic materials, quasi one-dimensional (1D) organometallic nanowires [4-8] are proposed as typical and important candidates[4-6] for their specific quasi 1D and layered geometries.

A series of experiments about organometallic complexes such as vanadium-benzene cluster $((V)_n(Bz)_m)$ with large $m$ and $n$ were reported[9-11]. And it was predicted exciting magnetic properties inside the vanadium-benzene nanowires (VBNW), such as the half-metal property[4,5]. Because of various combinations between metal and organic decker, the ground states can be quite different including ferromagnetic (FM), paramagnetic (PM), and antiferromagnetic (AFM) ones. Based on the experimental progresses, theoretic works also began to propose similar spintronic structures[12-17] or replacing benzene decker by other organic deckers[18-22].

Due to double exchange or charge transfer dependent on the specific choice of organic deckers,[5,6,22] metal's electron configurations and mechanisms of the origination of the magnetisms are quite different. These novel one-dimensional metal-organic nanowires are potential for future molecular spintronic applications, such as half metal,[4-6] negative differential resistance,[18] and spin filters.[4,6,18] Moreover, due to the variant exchange and charge-transfer properties between metal and deckers, the external factors can influence the interaction of metal and deckers and last affect the magnetic properties, revealing that the potential applications of spintronic sensors. Thus, magnetic properties of vanadium-benzene nanowires (VBNWs) affected by NO and $NO_2$ are investigated by first-principle calculations, and the results indicate that VBNW can be a molecular spintronic sensor to detect two kinds of gas molecules.

## 2. Theoretical methods and models

All the calculations were performed by a spin-polarized density functional theory (DFT) method within generalized gradient approximation (GGA) with PW91 functional[23], as implemented in Vienna *ab-initio* simulation package (VASP)[24,25]. Interactions between ions and electrons are described by the frozen core projector augmented wave method[26,27]. The cutoff energy of the plane-wave basis set is 500 eV, and a *k*-mesh of 1×1×45 within Monkhorst-Pack *k*-point scheme is adopted[28]. In the structural optimization, the convergence of Hellmann-Feynman forces is set to be 0.001 eV/Å per atom. For the tetragonal structures, a supercell has the dimension of $15 \times 15 \times c$ Å$^3$, where *c* is the periodic length along the axial direction (Figure 1). For the hexagonal structures of VBNW with $D_{6h}$ symmetry, a similar supercell with 120 degree is used to guarantee $D_{6h}$ symmetry. During the structural optimization, the periodic length *c* and all the atom positions are relaxed.

Figure 1 shows the top-view atomic structures of VBNW combining with NO and $NO_2$ gas molecules around the vanadium atoms. Since the structure with only one molecule adsorbing on the VBNW is strongly distorted and quite unstable, herein we only consider the structures adsorbed two or three molecules. VBNW-2M(M=NO and $NO_2$) and VBNW-3M are defined for two and three molecule absorbed on the vanadium atoms of VBNW. Due to the relative positions of the molecules adsorbed on the VBNW, there are two typical geometric configurations (or isomers), denoted as (I) and (II) in Figure 1.

## 3. Results and discussion

Table 1 summarizes the electronic data of VBNW and VBNW-2M (or -3M) (I) (or (II)), including the lattice constant, ground states, band gaps, Bader charge of vanadium atom, magnetic moments, energy difference between AFM and FM states, and binding energies. The binding energy ($E_b$) is defined as $E_b = E_{VBNM-M} - (E_{VBNW} + E_M)$, where $E_{VBNW}$, $E_M$, and $E_{VBNM-M}$ are the energies of VBNW, molecules, and VBNW adsorbing molecules. The pristine VBNW is half-metal, and its electronic and magnetic properties are tuned by adsorbing various gas molecules. Based on the definition of half metal, a half-metal material is still ferromagnetic (FM) as well. The magnetic moment is about 0.80 $\mu_B$, in agreement with previous report[5].

To well understand electronic and magnetic properties of VBNW with gas molecules, the band structures and partial density of states (PDOS) are plotted in Figure 2 and Figure 3. Fermi level is set to zero. Since the electronic structures of two configuration (I) and (II) are very close (Table 1 and Figure 2), thus only one configuration is considered for the PDOS plots.

When two NO molecules adsorbed onto the vanadium atom, both VBNW-2NO (I) and (II) are ferromagnetic metals (FM-metals). Meanwhile, the binding energies are 0.15 eV. With further adsorption of NO, the systems of VBNW-3NO (I) and (II) become paramagnetic (PM) semiconductor with a zero magnetic moment, and the binding energy increases to -3.4 eV, and the band gap is about 0.65 eV. Thus there is a (half-metal)-(FM-metal)-(PM-semiconductor) transition depending on the amount of NO molecules.

For $NO_2$ series, they show different behaviors. In either VBNW-$2NO_2$ (I) or (II), the band structures reveal that they possess a conductor of up-spin part and a semicondutor of down-spin part, that is, they are half-metals. The magnetic moment increases to 3 $\mu_B$ in comparison to 0.80 $\mu_B$ in pristine VBNW. After the addition adsorption of $NO_2$, VBNW-$3NO_2$ (I) and (II) become FM semiconductors. Their binding energies increase and the magnetic moments decrease, similar to VBNE-3NO cases.

When gas molecules adsorb on the vanadium atom, the lattice constant $c$ of the structure is enlarged, as shown in Table 1. For example, in VBNM-2NO cases, $c$ is 4.32Å, which is about 1Å larger than that in VBNW structure. The binding between vanadium atom and benzene decker is weaken after the adsorption of NO and $NO_2$ molecules, while the coupling between vanadium and nitrogen is strengthened.

Moreover, charge transfer between metal and molecules can be observed from Bader charge analysis as shown in Table 1. For NO series, Bader charge of vanadium decreases from 3.45 $e$ in VBNW to 3.35 $e$ in VBNW-2NO and to 3.07 $e$ in VBNW-3NO structures. Due to different electron affinity of $NO_2$ molecules, Bader charge of vanadium increased to 3.47 $e$ in VBNW-$2NO_2$ and decrease to 3.15 $e$ in VBNW-$3NO_2$. The vanadium atom loses more electrons with the increase of gas molecules. Since the charge transfers mainly occur in the $d$ orbitals, it's necessary to further investigate the orbitals' details. Thus partial DOS of 3$d$ orbital components are plotted in Figure 4.

In the pristine VBNW, the occupied states near Fermi level are mainly contributed by $d_{z2}$, $d_{xz}$ and $d_{yz}$ components.[4,5] In VBNW-2M structures, there is no degeneracy among five components of 3$d$ orbitals, as shown in Figure 4 (A) and (C). However in VBNW-3M structures (Figure 4 (B) and (D)), five components of 3$d$ orbitals of vanadium atom divide into three groups: (1) two folds degenerated $d_{yz}$ and $d_{xy}$, (2) $d_{xz}$ and $d_{x2-y2}$, and (3) $d_{z2}$.

With comprehensive analysis, it is found that in VBNWS-2M structures, the net charge transferred to NO and NO$_2$ are 1.37 $e$ and 1.64 $e$ respectively, while for VBNWS-3M cases, both transferred charges are 2.00 $e$, revealing that the extra coupling between the vanadium and NO (or NO$_2$) molecule affect the electron occupation in up/down-spin channels and the shapes of the spin densities (Figure 5) due to the charges transfer induced by the adsorption of molecules.

## 4. Conclusion

In summary, we perform spin-polarized DFT calculation on the electronic and magnetic properties of the quasi one-dimensional vanadium-benzene nanowires (VBNW) with the adsorption of two kinds of molecules. With the increase amount of NO adsorption on VBNW, phase transitions from half metal to ferromagnetic metal and last to paramagnetic semiconductor can be predicted theoretically. When adsorbing NO$_2$, half metallic property can be enhanced at first and decreased later with the increase quantity of NO$_2$ molecules. Due to the variant vanadium-molecule couplings, the electronic and magnetic properties of VBNW can be sensitive and selective to the external gas molecules including NO and NO$_2$, revealing the potential applications in molecular spintronic sensors.

### Acknowledgements

This work was financially supported by ….

### References

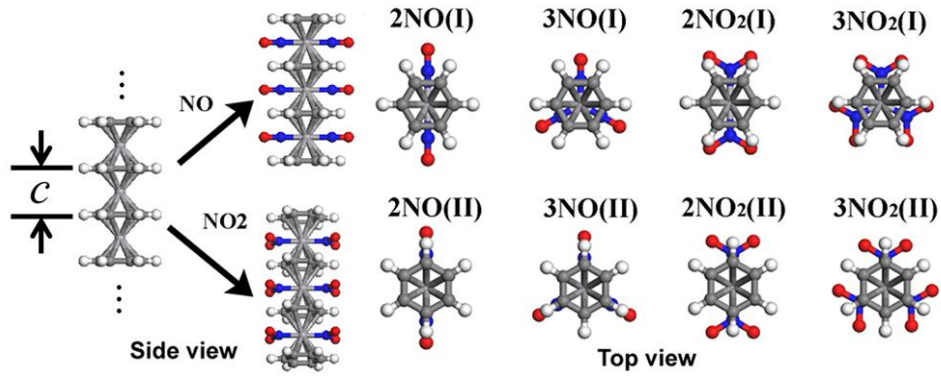

**Figure 1.** Atomic structure of VBNW with adsorptions of NO and $NO_2$ molecules. The silver, red, blue and white color representative the vanadium, oxygen, nitrogen and hydrogen atoms respectively. The number I and II represent two kinds of isomers.

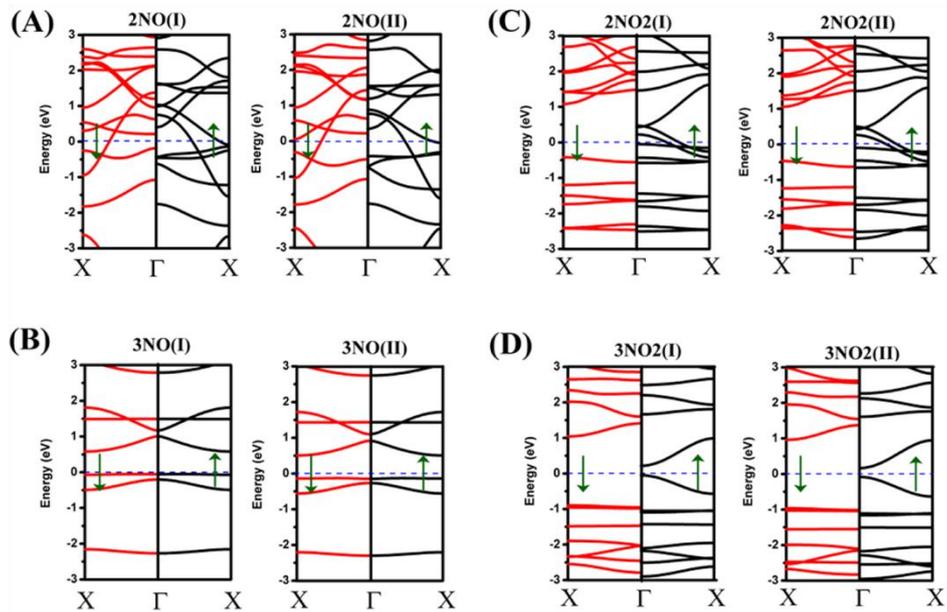

**Figure 2.** Band structures of VBNW with adsorptions of (A) 2NO, (B)3NO, (C)2$NO_2$ and (D)3$NO_2$. The up and down spin channels are in black and red and denoted by the green arrow. Fermi level is set to zero in each band structure.

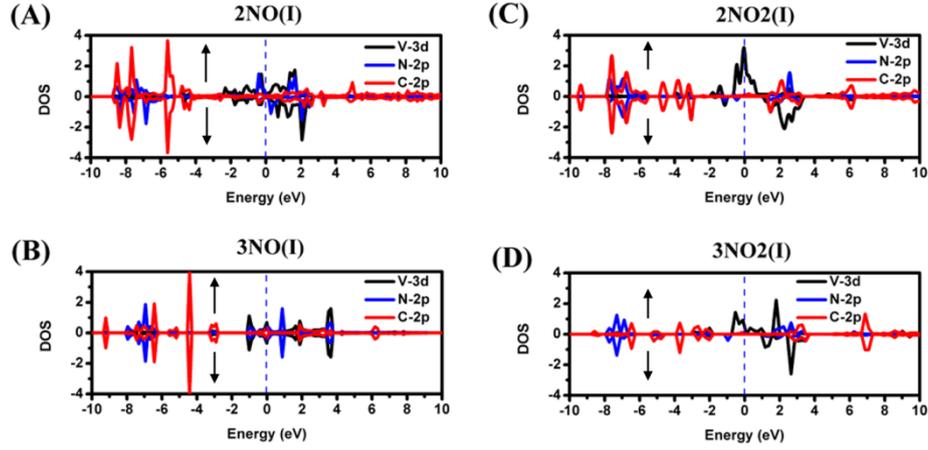

**Figure 3.** Partial density of states plot for VBNW- (A) 2NO(I), (B) 3NO(I), (C) 2NO2(I), and (D) 3NO2(I) structures. The 3d orbitals of V, 2p orbitals of N and C are considered. Fermi level is set to zero.

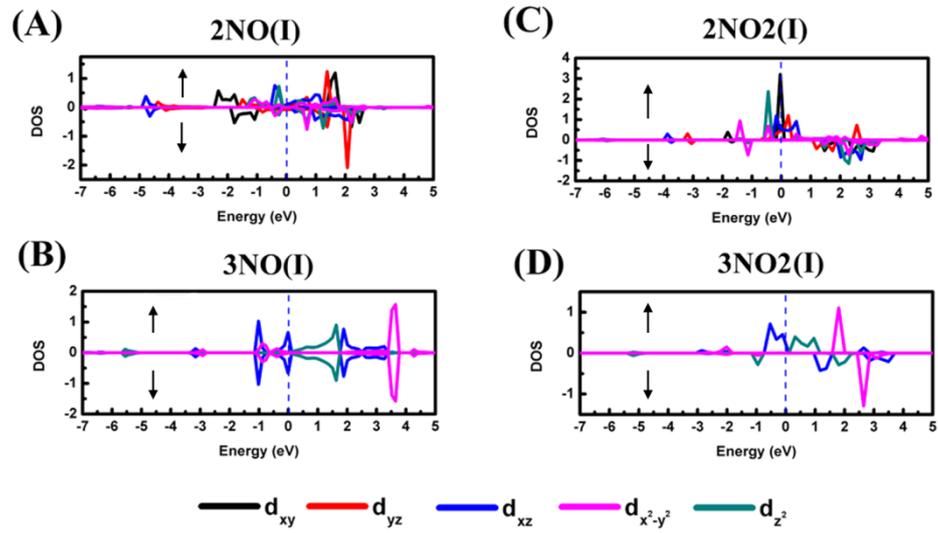

**Figure 4.** Partial density of states plot of 3$d$ orbitals for VBNW- (A) 2NO(I), (B) 3NO(I), (C) 2NO$_2$(I), and (D) 3NO$_2$(I).

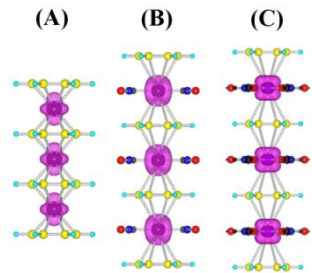

**Figure 5.** Spin densities of (A)VBNW, (B) VBNW-2NO$_2$(I) and (C) VBNW-3NO$_2$(I).

**Table 1.** The lattice constant ($c$), ground states (GS), band gaps (for insulator cases),

Bader charge of vanadium atom (*e*), magnetic moment (Mag); energy difference between FM and AFM states (ΔE); and binding energy $E_b = E_{VBNM-M} - (E_{VBNW} + E_M)$, where $E_{VBNW}$, $E_M$, and $E_{VBNM-M}$ are the energies of VBNW, molecules, and VBNW adsorbing molecules.

| Structures | c(Å) | GS | Band gap(*eV*) | Bader charge(*e*) | Mag ($\mu_B$) | ΔE (*eV*) | $E_b$ (*eV*) |
|---|---|---|---|---|---|---|---|
| VBNM | 3.34 | half-metal | - | 3.45 | 0.80 | -0.08 | / |
| VBNM-2NO(I) | 4.32 | FM-metal | / | 3.35 | 2.29 | -0.26 | -0.15 |
| VBNM-2NO(II) | 4.32 | FM-metal | / | 3.35 | 2.29 | -0.02 | -0.13 |
| VBNM-3NO(I) | 5.14 | PM-semiconductor | 0.65 | 3.07 | 0.00 | 0.00 | -3.42 |
| VBNM-3NO(II) | 5.14 | PM-semiconductor | 0.65 | 3.07 | 0.00 | 0.00 | -3.40 |
| VBNM-2NO$_2$(I) | 5.12 | half-metal | / | 3.47 | 3.00 | -0.32 | -0.91 |
| VBNM-2NO$_2$(II) | 5.12 | half-metal | / | 3.47 | 3.00 | -0.26 | -0.83 |
| VBNM-3NO$_2$(I) | 5.46 | FM-semiconductor | 0.24 | 3.15 | 2.00 | -1.02 | -3.87 |
| VBNM-3NO$_2$(II) | 5.46 | FM-semiconductor | 0.24 | 3.15 | 2.00 | -1.29 | -3.86 |